\documentclass[11pt,a4paper]{article}

\usepackage{amsmath,amssymb,bm}
\usepackage{slashed}
\usepackage{hyperref}
\usepackage{microtype}
\usepackage[margin=2.4cm]{geometry}
\usepackage{booktabs}
\usepackage{graphicx}
\usepackage{xcolor}
\usepackage{placeins}
\hypersetup{hidelinks}

\newcommand{\dd}{\mathrm{d}}
\newcommand{\ii}{\mathrm{i}}
\newcommand{\Tr}{\operatorname{Tr}}

\begin{document}

\title{Fermionic Casimir memory from an anisotropic gravitational pulse}

\author{
Celio R. Muniz\thanks{Corresponding author:
\href{mailto:celio.muniz@uece.br}{celio.muniz@uece.br}}\\
\small Faculdade de Educa\c{c}\~ao, Ci\^encias e Letras de Iguatu,\\
\small Universidade Estadual do Cear\'a,
Av. D\'ario Rabelo, s/n, Iguatu, CE, 63500-000, Brazil
\and
Herondy F. Santana Mota\thanks{
\href{mailto:hmota@fisica.ufpb.br}{hmota@fisica.ufpb.br}}\\
\small Departamento de F\'isica, Universidade Federal da Para\'iba,\\
\small 58059-970, Caixa Postal 5008, Jo\~ao Pessoa, PB, Brazil
}

\date{\today}

\maketitle

\begin{abstract}
We investigate the late-time energy deposited by a weak, spatially homogeneous anisotropic gravitational pulse in a massless Dirac field confined between two static parallel plates. The field satisfies MIT bag boundary conditions, while the prescribed geometry approaches Minkowski spacetime in both asymptotic time regions. Using a tetrad formulation, we show that the linear spin-connection contribution cancels for the traceless anisotropy, leaving a coupling to the shear component of the Dirac stress tensor. The pulse produces real fermion--antifermion pairs, whose deposited energy is evaluated through the corresponding Bogoliubov coefficients. Subtracting the production in unbounded space isolates a finite boundary-dependent contribution governed by the pulse duration relative to the characteristic spectral timescale set by the MIT gap. This contribution is negative in the adiabatic regime, where confinement suppresses excitations relative to the continuous free spectrum, and approaches zero from above in the impulsive limit. The late-time force on the plates is obtained separately from the normal stress $T^{zz}$ of the produced pairs, evaluated at fixed occupation numbers, with the field and the pulse present on both sides of each plate. The resulting pressure correction is the difference between the interior and exterior normal stresses. It reinforces the attractive Casimir force for both slow and fast pulses and reduces it in an intermediate window, so that it exhibits two sign changes. Unlike the bosonic vacuum-polarization memory previously obtained for the same class of backgrounds, the present effect is a dissipative fermionic memory carried by gravitationally created pairs. Here, memory denotes the persistent late-time modification of the quantum state after the pulse has vanished.
\end{abstract}

\section{Introduction}

The Casimir effect provides a well-established setting in which boundary conditions convert quantum-vacuum fluctuations into observable shifts of energy and stress~\cite{Casimir1948,Bordag2009}. Its interplay with gravitation is conceptually important because both the definition of vacuum and the spectral properties of confined quantum fields depend on the spacetime geometry. A substantial literature has therefore examined Casimir energies and forces in weak gravitational fields, accelerated frames, rotating backgrounds, and alternative theories of gravity~\cite{Sorge2005,Sorge2009,Sorge2019,FullingEtAl2007,BimonteEtAl2006}. In particular, gravitational corrections have been investigated in Ho\v{r}ava--Lifshitz backgrounds, near slowly rotating sources, in the presence of a cosmological constant and quintessence, and through explicit mode-sum analyses of weak-field cavities~\cite{MunizBezerraCunha2013,BezerraMotaMuniz2014,MunizBezerraCunha2015,BezerraEtAl2017,LimaAlencarMunizLandim2019}. These studies also show that the existence and magnitude of a gravitational correction can depend sensitively on the choice of coordinates, boundary conditions, perturbative order, and operational definition of the Casimir energy. A recent review summarizes several aspects of this gravitational Casimir program~\cite{BezerraEtAl2024}.

Most of these investigations concern stationary backgrounds and the corresponding modification of vacuum polarization or zero-point energy. Time-dependent geometries open a physically different channel: they can convert vacuum fluctuations into real excitations~\cite{BirrellDavies,ParkerToms}. In the conventional dynamical Casimir effect, particle production is driven by moving boundaries or by externally modulated boundary conditions, whereas in quantum field theory in curved spacetime the driving agent may be the geometry itself. When the background becomes stationary in the asymptotic past and future, unambiguous in- and out-particle states can be introduced and related by a Bogoliubov transformation. A nonvanishing mixing between positive- and negative-frequency modes then signals vacuum instability and particle creation.

Fermionic realizations are especially interesting because their response is shaped by spinor structure, Pauli statistics, and confining boundary conditions with no direct scalar analogue. Fermion production by anisotropic cosmological evolution has been studied in unbounded geometries, including the production of massless particles in Bianchi-I backgrounds~\cite{Bhoonah2019}. Fermionic dynamical Casimir effects have also been investigated for moving bag boundaries in lower-dimensional systems~\cite{FoscoHansen2022,FoscoHansen2023,FoscoHansen2024}, while functional, spectral, and mode-sum approaches have clarified the role of fermionic confinement in static Casimir problems~\cite{SundbergJaffe2004,FoscoLosada2008,Elizalde2003,MandlechaGavai2022}. Nevertheless, the combined effect of a transient anisotropic gravitational background and static MIT bag boundaries on the late-time energy of gravitationally produced Dirac pairs remains comparatively unexplored.

Sorge recently investigated the response of a static Casimir cavity to a weak Bianchi-I pulse~\cite{Sorge2023}. Although the geometry returns asymptotically to Minkowski spacetime, the pulse leaves a persistent modification of the confined quantum state. His calculation concerns vacuum polarization, obtained from the real part of Schwinger's proper-time effective action, and the resulting shift in the zero-point energy of confined massless scalar and electromagnetic fields. It therefore establishes an important bosonic realization of gravitational Casimir memory, but does not determine the dissipative fermionic response. In the present case, the memory is instead associated with vacuum decay and real fermion--antifermion production, encoded in the Bogoliubov coefficients and related to the absorptive sector of the in--out effective action~\cite{Schwinger1951}.

In this work, we consider a massless Dirac field in a $(3+1)$-dimensional slab bounded by two static parallel plates satisfying MIT bag conditions and subjected to a weak, spatially homogeneous anisotropic gravitational pulse. We calculate the late-time energy carried by the produced fermion--antifermion pairs and isolate its boundary-dependent part by subtracting the corresponding production in unbounded space at the same pulse parameters. We refer to the resulting finite remnant as \emph{fermionic Casimir memory}. The response is controlled by the competition between the pulse bandwidth and the gap of the half-integer MIT spectrum. We then compute the late-time force on the plates directly from the normal stress of the produced pairs at fixed occupation numbers, balancing the interior against the exterior. This yields a nonmonotonic correction to the Casimir pressure, including regimes in which the usual attraction is either reinforced or reduced. Here, ``memory'' denotes the persistent late-time modification of the quantum state after the prescribed pulse has vanished and should not be confused with the classical displacement memory of a propagating gravitational wave.

The paper is organized as follows. In Sec.~2, we introduce the spatially homogeneous anisotropic pulse, formulate the massless Dirac problem in the slab, and impose the MIT bag boundary conditions, establishing the longitudinal spectrum and the associated selection rule. Section~3 derives the Bogoliubov coefficients and the late-time energy deposited in fermion--antifermion pairs. In Sec.~4, we perform the subtraction of the corresponding unbounded-space contribution and obtain the finite dimensionless response function governing the fermionic Casimir memory. Section~5 analyzes the adiabatic and impulsive regimes of the energy response, and Sec.~6 derives the late-time correction to the Casimir pressure from the normal stress of the produced pairs. Our main findings and possible extensions are discussed in Sec.~7. Technical details of the spin trace, normalization, and longitudinal MIT-mode overlap are collected in Appendices~A and~B.

We use units $\hbar=c=1$ and metric signature $(+---)$. Greek indices refer to the coordinate basis, whereas hatted Latin indices label the local Lorentz frame; accordingly, $\gamma^\mu(x)$ are curved-space matrices and $\gamma^{\hat a}$ are constant tangent-space Dirac matrices.

\section{Dirac field in the pulsed cavity}

The plates are thin, located at $z=0$ and $z=L$, and impose MIT conditions on both of their faces. The Dirac field exists in the interior slab $0<z<L$ and in the two exterior half-spaces $z<0$ and $z>L$. The pulse is spatially homogeneous and therefore acts in all three regions. This is the configuration in which the standard static fermionic Casimir pressure is defined, since the bulk contributions on the two sides of each plate cancel; it will matter in Sec.~6, where the force on the plates is computed. Everywhere outside the plates we consider
\begin{equation}
 \dd s^2=\dd t^2-[1+h(t)]\dd x^2-[1-h(t)]\dd y^2-\dd z^2,
 \label{metric}
\end{equation}
with a weak Gaussian pulse
\begin{equation}
 h(t)=H e^{-\sigma^2t^2},\qquad |H|\ll1.
 \label{pulse}
\end{equation}
The duration is of order $\tau=1/\sigma$. The metric pulse is spatially homogeneous and describes a weak,
time-dependent anisotropic deformation of Bianchi-I type, treated here
as a prescribed classical gravitational background. It represents a
transient shear that vanishes in both asymptotic time regions.
Accordingly, $H$ denotes the pulse amplitude, while $\sigma^{-1}$
sets its characteristic duration. Since $h(t)\to0$ as
$t\to\pm\infty$, the in and out particle concepts are unambiguous.

We stress that the homogeneous Gaussian shear in Eq.~\eqref{metric} is
not a vacuum solution of the Einstein equations: already at linear order
its Ricci tensor contains terms proportional to $\ddot h$, so it must be
regarded as sourced by an anisotropic stress or simply as a prescribed
external background. A gravitational wave propagating along $z$ has the
same tensor structure, but with $h(t-z)$; replacing it by $h(t)$ across
the cavity requires a wavelength much larger than $L$, i.e.\
$\sigma L\ll1$. The interpretation as a local approximation to a
propagating wave is therefore restricted to this long-wavelength
regime, as also assumed in Ref.~\cite{Sorge2023}. Moreover, an exact plane wave produces no particles in
unbounded space~\cite{Gibbons1975}, so the free-space production used
below as a reference is a property of the homogeneous pulse and not of
a propagating wave. The intermediate regime $\sigma L\sim1$--$10$,
where the effects found below are largest, should accordingly be
understood within the prescribed-background model.
The massless Dirac action is
\begin{equation}
 S_D=\int\dd^4x\sqrt{-g}\,\bar\psi\,\ii\gamma^\mu(x)\nabla_\mu\psi,
 \label{action}
\end{equation}
where $\gamma^\mu=e^\mu{}_{\hat a}\gamma^{\hat a}$. A convenient orthonormal coframe is
\begin{equation}
 \vartheta^{\hat0}=\dd t,\quad
 \vartheta^{\hat1}=\sqrt{1+h}\,\dd x,\quad
 \vartheta^{\hat2}=\sqrt{1-h}\,\dd y,\quad
 \vartheta^{\hat3}=\dd z.
 \label{tetrad}
\end{equation}

It is useful to display explicitly how the linearized Dirac operator follows
from Eq.~\eqref{tetrad}. Introduce the directional scale factors
\begin{equation}
 a_x(t)=\sqrt{1+h(t)},\qquad a_y(t)=\sqrt{1-h(t)}.
 \label{scale_factors}
\end{equation}
The inverse tetrad gives
\begin{equation}
 \gamma^t=\gamma^{\hat0},\qquad
 \gamma^x=\frac{\gamma^{\hat1}}{a_x},\qquad
 \gamma^y=\frac{\gamma^{\hat2}}{a_y},\qquad
 \gamma^z=\gamma^{\hat3}.
 \label{curved_gamma_matrices}
\end{equation}
For this diagonal Bianchi-I tetrad, the complete massless Dirac operator is
\begin{align}
 \ii\gamma^\mu\nabla_\mu
 ={}&\ii\gamma^{\hat0}
 \left[\partial_t+\frac12
 \left(\frac{\dot a_x}{a_x}+\frac{\dot a_y}{a_y}\right)\right]
 +\frac{\ii\gamma^{\hat1}}{a_x}\partial_x
 +\frac{\ii\gamma^{\hat2}}{a_y}\partial_y
 +\ii\gamma^{\hat3}\partial_z .
 \label{full_Dirac_operator}
\end{align}
For the compensated anisotropy in Eq.~\eqref{metric},
\begin{equation}
 \frac{\dot a_x}{a_x}+\frac{\dot a_y}{a_y}
 =-\frac{h\dot h}{1-h^2}=O(H^2),
 \qquad
 \sqrt{-g}=\sqrt{1-h^2}=1+O(H^2).
 \label{spincancel}
\end{equation}
Moreover,
\begin{equation}
 a_x^{-1}=1-\frac{h}{2}+O(H^2),\qquad
 a_y^{-1}=1+\frac{h}{2}+O(H^2).
 \label{inverse_scale_expansion}
\end{equation}
Consequently, to linear order in the pulse amplitude,
\begin{equation}
 \ii\gamma^\mu\nabla_\mu
 =\ii\gamma^{\hat\mu}\partial_\mu+\delta D+O(H^2),
 \qquad
 \delta D=-\frac{\ii h(t)}{2}\gamma^{\hat1}\partial_x
 +\frac{\ii h(t)}{2}\gamma^{\hat2}\partial_y.
 \label{deltaD}
\end{equation}
This calculation also shows why no term proportional to $\dot h$ survives at
first order: the two transverse logarithmic derivatives cancel because the
deformation is traceless to linear order. This cancellation is specific to
the compensated deformation in Eq.~\eqref{metric}; it is not a generic
property of anisotropic backgrounds.

To connect Eq.~\eqref{deltaD} with the stress tensor, we use the symmetric flat-space expression valid in the asymptotic regions,
\begin{equation}
 T^{\mu\nu}=\frac{\ii}{4}\bar\psi
 \left(\gamma^\mu\overleftrightarrow{\partial^\nu}
 +\gamma^\nu\overleftrightarrow{\partial^\mu}\right)\psi.
 \label{stress}
\end{equation}
The linear variation $\delta S=-\tfrac12\int\dd^4x\,\delta g_{\mu\nu}T^{\mu\nu}$, together with $\delta g_{xx}=-h$ and $\delta g_{yy}=h$, gives
\begin{equation}
 S_{\mathrm{int}}=\frac12\int\dd^4x\,h(t)
 \left(T^{xx}-T^{yy}\right).
 \label{Sint}
\end{equation}
Thus the pulse probes the anisotropic stress of the Dirac field rather than its trace.

The MIT conditions are~\cite{Johnson1975,Sitenko2015,MandlechaGavai2022}
\begin{equation}
 (1+\ii n_\mu\gamma^\mu)\psi\big|_{\partial\mathcal M}=0,
 \label{MIT}
\end{equation}
with oppositely directed outward normals at the two plates. They make the Dirac Hamiltonian self-adjoint and enforce a vanishing normal current. A mode in the slab is a superposition of the $+k_z$ and $-k_z$ branches. Successive reflection at the two MIT walls gives $e^{2\ii k_zL}=-1$ in the massless case, and hence
\begin{equation}
 k_{z,n}=\frac{\pi}{L}\left(n+\frac12\right),\qquad
 E_{n\bm k}=\sqrt{\bm k^2+k_{z,n}^2},
 \label{spectrum}
\end{equation}
where $n=0,1,\ldots$ and $\bm k=(k_x,k_y)$. In particular, the confined spectrum has the gap $\pi/(2L)$.

The absence of translational invariance along the normal direction means
that ordinary longitudinal-momentum conservation cannot be invoked.
Nevertheless, the homogeneous shear vertex is diagonal in the
longitudinal MIT spectrum. Indeed, defining
\begin{equation}
K_z=-i\alpha^{\hat 3}\partial_z,
\qquad
O_\perp(\mathbf{k})=
k_x\alpha^{\hat 1}-k_y\alpha^{\hat 2},
\label{eq:longitudinal_vertex}
\end{equation}
one finds
\begin{equation}
\{K_z,O_\perp\}=0,
\qquad
[\gamma^{\hat 3},O_\perp]=0 .
\label{eq:longitudinal_algebra}
\end{equation}
The second relation implies that $O_\perp$ commutes with the MIT
projectors and maps the boundary domain into itself. This alone does not
fix the derivative boundary data required by the second-order operator
$K_z^2=-\partial_z^2$. The argument is completed in
\ref{app:longitudinal_overlap} by working with the self-adjoint
first-order Hamiltonian $\mathcal H_{\bm k}=H_\perp+K_z$ at fixed
transverse momentum: $O_\perp V_{ms'}$ lies in the domain of
$\mathcal H_{\bm k}^2$, the corresponding Green boundary form vanishes,
and $O_\perp V_{ms'}$ is an eigenfunction of $\mathcal H_{\bm k}^2$ with
eigenvalue $E_{m\bm k}^2$. Since different longitudinal indices $n$
correspond to different eigenvalues, orthogonality between different
eigenspaces gives
\begin{equation}
\int_0^L dz\,
U_{ns}^{\dagger}(z)O_\perp(\mathbf{k})V_{ms'}(z)
=
\delta_{nm}\,\mathcal V_{nss'}.
\label{eq:longitudinal_selection}
\end{equation}
Thus the pulse creates a pair within the same longitudinal MIT level.
The walls can exchange longitudinal momentum, but they do not generate
an independent double sum over $n$ and $m$ for this spatially
homogeneous shear perturbation. The proof and an explicit check with the
complete standing-wave MIT modes are given in
\ref{app:longitudinal_overlap}. The standard massless MIT
spectrum contains no additional surface-bound branch.

\section{Pair production}

For fixed $n$ and $\bm k$, the linear interaction Hamiltonian contains the spinor vertex
\begin{equation}
\mathcal{O}_{\perp}(\mathbf{k})
=
k_x\alpha^{\hat 1}-k_y\alpha^{\hat 2}.
\label{vertex}
\end{equation}
Spatial homogeneity imposes opposite transverse momenta on the created
particle and antiparticle, while Eq.~\eqref{eq:longitudinal_selection}
enforces equality of their longitudinal mode indices. Inserting the
asymptotic MIT-mode expansion into Eq.~\eqref{Sint}, the first-order
vacuum-to-pair amplitude is
\begin{align}
\beta_{nss'}(\mathbf{k})
={}&-\ii\int_{-\infty}^{\infty}\dd t\,
 \langle n,\mathbf{k},s; n,-\mathbf{k},s'|
 H_{\mathrm{int}}(t)|0\rangle
 \nonumber\\
={}&-\frac{\ii}{2}\int_{-\infty}^{\infty}\dd t\,
 h(t)e^{2\ii E_{n\mathbf{k}}t}
 \int_0^L\dd z\,
 U_{ns}^{\dagger}(z)\mathcal{O}_{\perp}(\mathbf{k})V_{ns'}(z)
 \nonumber\\
={}&-\frac{\ii}{2}\widetilde h(2E_{n\mathbf{k}})
 \mathcal V_{nss'} .
\label{beta}
\end{align}
Here $\widetilde h(\Omega)=\int\dd t\,h(t)e^{\ii\Omega t}$. For Eq.~\eqref{pulse},
\begin{equation}
 |\widetilde h(2E)|^2=\frac{\pi H^2}{\sigma^2}e^{-2E^2/\sigma^2}.
 \label{fourier}
\end{equation}

The spin sum may be evaluated without choosing a particular gamma-matrix representation. Defining the covariant vertex $\mathcal W=k_x\gamma^{\hat1}-k_y\gamma^{\hat2}$, the completeness relations for opposite particle and antiparticle momenta reduce it to
\begin{align}
 \sum_{s,s'}|\bar u_s\mathcal Wv_{s'}|^2
 &=\Tr\left[(\slashed p)\mathcal W(\slashed q)\overline{\mathcal W}\right]
 \nonumber\\
 &=8\left[E^2k_\perp^2-(k_x^2-k_y^2)^2\right],
 \label{spintrace}
\end{align}
where $q=(E,-\bm p)$ and the normalization factors entering Eq.~\eqref{beta} are understood. Writing $k_x=k_\perp\cos\phi$ and $k_y=k_\perp\sin\phi$, the only angular average required is $\langle\cos^2(2\phi)\rangle=1/2$. Hence
\begin{equation}
 \left\langle\sum_{s,s'}|\bar u_s\mathcal W v_{s'}|^2\right\rangle_\phi
 =8\left(E^2k_\perp^2-\frac{k_\perp^4}{2}\right).
 \label{spinsum}
\end{equation}
Multiplying $|\beta|^2$ by the pair energy $2E$ gives the deposited energy per unit plate area,
\begin{align}
 \frac{\Delta E_D(L)}{A}
 ={}&\frac{\pi H^2}{\sigma^2}\sum_{n=0}^{\infty}
 \int\frac{\dd^2\bm k}{(2\pi)^2}e^{-2E_{n\bm k}^2/\sigma^2}
 \nonumber\\
 &\times\left(\frac{k_\perp^4}{2E_{n\bm k}}
 +\frac{k_{z,n}^2k_\perp^2}{E_{n\bm k}}\right).
 \label{cavityenergy}
\end{align}
The factor 8 in Eq.~\eqref{spinsum} already includes the particle and antiparticle spin sums; no additional spin degeneracy is required. Equation~\eqref{cavityenergy} is finite and non-negative. It is not yet a Casimir contribution, because the pulse also creates pairs in the absence of plates.

\section{Casimir subtraction and response function}

We define the fermionic Casimir memory by subtracting the unbounded-space result at the same pulse parameters,
\begin{equation}
 \frac{\Delta E_{\mathrm C}}{A}
 \equiv\frac{\Delta E_D(L)}{A}-\frac{\Delta E_D(\infty)}{A}.
 \label{subtraction}
\end{equation}
The continuum replacement is
\begin{equation}
 \sum_{n=0}^{\infty}\longrightarrow
 \frac{L}{\pi}\int_0^\infty\dd k_z.
 \label{continuum}
\end{equation}
To make this reduction explicit, write
$\dd^2\bm k=2\pi k_\perp\dd k_\perp$ and introduce
$u=(k_\perp^2+k_z^2)^{1/2}$, so that
$k_\perp\dd k_\perp=u\dd u$. For each fixed $k_z$, the transverse
integral in Eq.~\eqref{cavityenergy} becomes
\begin{align}
&\int\frac{\dd^2\bm k}{(2\pi)^2}
 e^{-2(k_\perp^2+k_z^2)/\sigma^2}
 \left(\frac{k_\perp^4}{2u}
 +\frac{k_z^2k_\perp^2}{u}\right)
 \nonumber\\
&\hspace{2cm}=
 \frac{1}{4\pi}\int_{k_z}^{\infty}\dd u\,
 (u^4-k_z^4)e^{-2u^2/\sigma^2}.
 \label{radial_reduction}
\end{align}
Therefore,
\begin{equation}
 \frac{\Delta E_D(L)}{A}
 =\frac{H^2}{4\sigma^2}
 \sum_{n=0}^{\infty}\int_{k_{z,n}}^{\infty}\dd u\,
 (u^4-k_{z,n}^4)e^{-2u^2/\sigma^2}.
 \label{cavityenergy_radial}
\end{equation}
With the dimensionless variables $y=uL$, $b=k_zL$, and
\begin{equation}
 \xi\equiv\sigma L
 \label{xi}
\end{equation}
the result assumes the scaling form
\begin{equation}
 \frac{\Delta E_{\mathrm C}}{A}=\frac{H^2}{L^3}F_D(\xi).
 \label{mainresult}
\end{equation}
The dimensionless response is
\begin{equation}
 F_D(\xi)=\frac{1}{4\xi^2}
 \left[\sum_{n=0}^{\infty}K\left(\pi(n+\tfrac12);\xi\right)
 -\frac{\xi^6}{10\pi}\right],
 \label{FD}
\end{equation}
where
\begin{equation}
 K(b;\xi)=\int_b^\infty\dd y\,(y^4-b^4)e^{-2y^2/\xi^2}.
 \label{kernel}
\end{equation}
For numerical work it is useful to write
\begin{align}
 K(b;\xi)={}&\frac12\left(\frac{\xi^2}{2}\right)^{5/2}
 \Gamma\left(\frac52,\frac{2b^2}{\xi^2}\right)
 \nonumber\\
 &-\frac{b^4}{2}\left(\frac{\xi^2}{2}\right)^{1/2}
 \Gamma\left(\frac12,\frac{2b^2}{\xi^2}\right).
 \label{kernelgamma}
\end{align}
The last term in Eq.~\eqref{FD} is the exact continuum contribution. Indeed, exchanging the order of integration gives
\begin{align}
 \int_0^\infty\dd b\,K(b;\xi)
 &=\int_0^\infty\dd y\,e^{-2y^2/\xi^2}
 \int_0^y\dd b\,(y^4-b^4)
 \nonumber\\
 &=\frac45\int_0^\infty\dd y\,y^5e^{-2y^2/\xi^2}
 =\frac{\xi^6}{10},
 \label{continuumderivation}
\end{align}
so that
\begin{equation}
 \frac1\pi\int_0^\infty\dd b\,K(b;\xi)=\frac{\xi^6}{10\pi}.
 \label{continuumintegral}
\end{equation}
No ultraviolet regulator is required for the produced energy: the Fourier profile of the physical pulse supplies the Gaussian factor in Eq.~\eqref{cavityenergy}. There is likewise no infrared divergence, because the integrand vanishes with sufficient powers of transverse momentum. This differs from the regularization of the static zero-point energy, for which heat-kernel or zeta-function methods are standard~\cite{Vassilevich2003,Bordag2009}. At large $n$, the discrete kernel is a polynomial times $\exp[-2\pi^2(n+1/2)^2/\xi^2]$ and is therefore exponentially convergent for every finite $\xi$. For large $\xi$, however, the two finite terms in Eq.~\eqref{FD} are close; Euler--Maclaurin or Poisson resummation is then numerically preferable to direct subtraction at machine precision.

Adding the standard static massless MIT contribution~\cite{Elizalde2003,MandlechaGavai2022} gives the sum of
the static Casimir energy and the boundary-dependent produced-pair
energy,
\begin{equation}
 \frac{E_{\mathrm{sub}}^D}{A}
 =-\frac{7\pi^2}{2880L^3}
 +\frac{H^2}{L^3}F_D(\xi)+O(H^4).
 \label{totalenergy}
\end{equation}
Both terms are bulk-subtracted. In particular, the second term is not the
total late-time energy stored inside the cavity, which is
$\Delta E_D(L)/A$ and is positive; it is the difference between that
energy and the energy deposited by the same pulse in a free region of
equal volume. The common factor $L^{-3}$ in Eq.~\eqref{totalenergy} is a
scaling statement at fixed $\xi$. At fixed physical pulse width,
$F_D(\sigma L)$ supplies additional $L$ dependence. As discussed in
Sec.~6, the $L$-derivative of Eq.~\eqref{totalenergy} at fixed pulse
parameters does not give the late-time force on the plates.
\section{Limiting regimes of the energy response}

In the adiabatic regime, $\xi\ll1$, the lowest dimensionless threshold is $b_0=\pi/2$. Hence every discrete term contains at least the factor $\exp[-2b_0^2/\xi^2]=\exp[-\pi^2/(2\xi^2)]$. The algebraic leading term comes entirely from the continuum subtraction in Eq.~\eqref{FD}, and therefore
\begin{equation}
 F_D(\xi)=-\frac{\xi^4}{40\pi}
 +O\left(e^{-\pi^2/(2\xi^2)}\right).
 \label{adiabatic}
\end{equation}
Although the energy created in either geometry is non-negative, their difference is negative: an adiabatic pulse can excite arbitrarily soft free-space modes but cannot bridge the MIT gap.

This negative sign does not represent the creation of negative-energy particles. It states that the plates reduce the positive energy deposited by the pulse relative to unbounded space. Nor is it, by itself, a statement about a local energy condition, which would require the full renormalized stress tensor rather than the global difference in Eq.~\eqref{subtraction}.

For $\xi\gg1$, the half-integer Euler--Maclaurin formula determines the small difference in Eq.~\eqref{FD}. In the form needed here,
\begin{equation}
 \sum_{n=0}^\infty f\left(n+\frac12\right)-\int_0^\infty f(x)\dd x
 =-\sum_{k=0}^\infty\frac{B_{k+1}(1/2)}{(k+1)!}f^{(k)}(0).
 \label{EMhalf}
\end{equation}
Near $b=0$ the kernel has the expansion
\begin{equation}
 K(b;\xi)=K(0;\xi)-A_0b^4+\frac45b^5
 -\frac{8}{21\xi^2}b^7+\cdots,
 \label{Ksmallb}
\end{equation}
where the $b^4$ term does not contribute to Eq.~\eqref{EMhalf}. In particular, the first and third derivatives vanish, while $K^{(5)}(0;\xi)=96$. Applying Eq.~\eqref{EMhalf} to $f(x)=K(\pi x;\xi)$ and using $B_6(1/2)=-31/1344$ gives
\begin{equation}
F_D(\xi)=\frac{31\pi^5}{40320\xi^2}
 +O(\xi^{-4}).
 \label{impulsive}
\end{equation}
Thus the Casimir part also vanishes for a pulse of vanishing integrated duration, now approaching zero from above. The response must consequently change sign between the slow and rapid regimes. Stable numerical evaluation of Eq.~\eqref{FD} gives
\begin{equation}
 \xi_0=1.98158\ldots,
 \label{zero}
\end{equation}
for the first zero. The response reaches a negative minimum $F_D=-0.0144024\ldots$ at $\xi=1.44159\ldots$ and a positive maximum $F_D=0.0382456\ldots$ at $\xi=3.09115\ldots$. These extrema refer to the boundary-dependent difference, not to the separately positive pair energies. In particular, the zero marks equality between the cavity and free-space deposited energies after spectral weighting by the pair energy.

The result should be distinguished from the memory effect reported in
Ref.~\cite{Sorge2023}, and the distinction is best stated in terms of the
observables involved. The quantity computed here is an in--in expectation
value: the late-time energy, and in Sec.~6 the normal stress, above their
out-vacuum values, in the state that evolves from the in vacuum. For a
free field whose geometry and boundaries return to their initial
configuration, the out vacuum is the vacuum of the same static cavity,
and these quantities are completely determined by the out-mode
occupations, e.g.\ $\sum_jN_jE_j$ for the energy. No separate
state-independent shift of the final vacuum energy can be added to them.
In Ref.~\cite{Sorge2023}, by contrast, the late-time shift of the energy
density of a confined massless scalar field with Dirichlet plates is
defined through the real part of Schwinger's proper-time effective
action, $\langle\epsilon_{\mathrm{Cas}}\rangle
=-(AL)^{-1}\lim_{t\to\infty}\partial_t\,\mathrm{Re}\,W$. In that
construction, the Bogoliubov weight $|\beta_p|^2$ enters the proper-time
trace multiplied by an off-shell kernel in which the mode index, the
frequency and the transverse momentum are integrated independently, and
the $L$-independent and spatially uniform contributions are removed by
counterterms. The resulting quantity is therefore not the in--in
expectation value of the energy in the out state. The difference is
already visible in the adiabatic limit. At fixed $L$, the in--in energy
and stress computed here vanish as $\sigma\to0$, as expected for a slow
perturbation of a gapped cavity [cf.\ Eqs.~\eqref{adiabatic} and
\eqref{eq:pressure_adiabatic}], whereas the shift obtained in
Ref.~\cite{Sorge2023} grows as $(\sigma L)^{-3}$ for $\sigma L\ll1$, a
behaviour also discussed there. The different field content and boundary
conditions additionally prevent a direct comparison between the response
functions.

\section{Late-time pressure on the plates}

\subsection{Why the energy derivative is not the force}

A natural first guess is to differentiate Eq.~\eqref{mainresult} with
respect to $L$ at fixed $H$ and $\sigma$. This does not give the force
exerted by the state produced in a given cavity, for two reasons.

First, changing $L$ changes the state prepared by the pulse, including the
occupation numbers $N_j$ of the out modes. The derivative therefore
compares two different preparation protocols. The mechanical force
exerted by a given late-time state follows instead from an adiabatic
virtual displacement of the plates, during which the occupations are
held fixed. Since $E_j=(\bm k^2+k_{z,n}^2)^{1/2}$ and
$\partial k_{z,n}/\partial L=-k_{z,n}/L$,
\begin{equation}
 P_{\mathrm{in}}
 =-\frac1A\sum_jN_j\frac{\partial E_j}{\partial L}
 =\frac{1}{AL}\sum_jN_j\frac{k_{z,j}^2}{E_j},
 \label{Pin_virtual}
\end{equation}
where $j$ runs over particle and antiparticle modes and transverse sums are
understood as integrals. By contrast,
\begin{equation}
 -\frac1A\frac{\dd}{\dd L}\sum_jN_j(L)E_j(L)
 =P_{\mathrm{in}}-\frac1A\sum_jE_j\frac{\dd N_j}{\dd L},
\end{equation}
and the last term is not a mechanical pressure.

Second, differentiating the subtracted term $-L\epsilon_{\mathrm{free}}$
in Eq.~\eqref{subtraction} contributes $+\epsilon_{\mathrm{free}}$, the
free-space energy density of the produced pairs. A population of pairs
outside the plates instead pushes on them with its normal stress
$p_{z,\mathrm{free}}$, and, as shown below, this stress differs from the
energy density because the produced gas is anisotropic.

\subsection{Pressure from the normal stress}

Equation~\eqref{Pin_virtual} coincides with the expectation value of the
normal stress at the walls. For a single normalized MIT mode one finds,
using the Dirac equation, $T^{zz}(z)=k_{z,n}^2/(E_{n\bm k}L)$ per unit
area, independent of $z$; the independence of $z$ follows from
$\partial_\mu T^{\mu z}=0$ for a stationary, transversely homogeneous
contribution. 

The late-time state also contains terms built from the anomalous
correlators $\langle a_jb_{j'}\rangle\propto\alpha\beta^*$, which
oscillate as $e^{-2\ii E_jt}$. Their contribution to $\langle
T^{zz}\rangle$ vanishes identically at linear order in $H$: the
transformation $H\to-H$ is equivalent to the reflection
$x\leftrightarrow y$, which leaves $T^{zz}$ and the plates invariant, so
$\langle T^{zz}\rangle$ is even in $H$ at all times. At order $H^2$ the
oscillatory terms carry phases $e^{-2\ii E_{n\bm k}t}$ integrated over a
continuous transverse spectrum. They dephase at late times and vanish
under time averaging. The late-time pressure discussed below is therefore
the time-averaged, stationary contribution from the occupation numbers.

Using the occupations implied by Eqs.~\eqref{beta}--\eqref{cavityenergy},
counting both members of each pair and repeating the radial reduction of
Eq.~\eqref{radial_reduction}, the interior normal stress becomes
\begin{equation}
 P_{\mathrm{in}}
 =\frac{H^2}{4\sigma^2L}\sum_{n=0}^{\infty}k_{z,n}^2
 \int_{k_{z,n}}^{\infty}\dd u\,
 \frac{u^4-k_{z,n}^4}{u^2}\,e^{-2u^2/\sigma^2}.
 \label{Pin_result}
\end{equation}
In each exterior half-space the same construction applies with a
continuous longitudinal spectrum. Since $T^{zz}$ is uniform there as
well, its value at the wall equals its bulk value, i.e.\ the normal stress
of the pairs produced in unbounded space. With Eq.~\eqref{continuum},
\begin{equation}
 p_{z,\mathrm{free}}
 =\frac{H^2}{4\pi\sigma^2}\int_0^\infty\dd k_z\,k_z^2
 \int_{k_z}^{\infty}\dd u\,\frac{u^4-k_z^4}{u^2}e^{-2u^2/\sigma^2}
 =\frac{H^2\sigma^4}{168\pi}.
 \label{pz_free}
\end{equation}
The corresponding energy density is
$\epsilon_{\mathrm{free}}=H^2\sigma^4/(40\pi)$, from
Eq.~\eqref{continuumintegral}. Hence
$p_{z,\mathrm{free}}=5\epsilon_{\mathrm{free}}/21$, which is smaller than
the isotropic value $\epsilon_{\mathrm{free}}/3$: the shear vertex
preferentially produces pairs with large transverse momentum. Tracelessness
of the massless stress tensor then gives
$p_{x,\mathrm{free}}=p_{y,\mathrm{free}}=8\epsilon_{\mathrm{free}}/21$.

The net memory-induced pressure, positive when it pushes the plates
apart, is the difference between the interior and exterior normal
stresses,
\begin{equation}
 \Delta P_{\mathrm{mem}}=P_{\mathrm{in}}-p_{z,\mathrm{free}}
 =\frac{H^2}{L^4}\,\mathcal P_D(\xi),
 \label{Pmem}
\end{equation}
with
\begin{equation}
 \mathcal P_D(\xi)=\frac{1}{4\xi^2}
 \left[\sum_{n=0}^{\infty}b_n^2\,J(b_n;\xi)
 -\frac{\xi^6}{42\pi}\right],
 \qquad b_n=\pi\left(n+\tfrac12\right),
 \label{PD}
\end{equation}
and
\begin{align}
 J(b;\xi)&=\int_b^\infty\dd y\,\frac{y^4-b^4}{y^2}\,e^{-2y^2/\xi^2}
 \nonumber\\
 &=\frac12\left(\frac{\xi^2}{2}\right)^{3/2}
 \Gamma\left(\frac32,\frac{2b^2}{\xi^2}\right)
 -\frac{b^4}{2}\left(\frac{2}{\xi^2}\right)^{1/2}
 \Gamma\left(-\frac12,\frac{2b^2}{\xi^2}\right).
 \label{Jkernel}
\end{align}
The last term in Eq.~\eqref{PD} is the exact continuum value,
$\pi^{-1}\int_0^\infty\dd b\,b^2J(b;\xi)=\xi^6/(42\pi)$, obtained by
exchanging the order of integration as in
Eq.~\eqref{continuumderivation}. If the field were absent outside the
plates, only $P_{\mathrm{in}}$ would remain; it is positive and grows as
$\sigma^4$ for short pulses, reflecting the pressure of an unbalanced gas
rather than a boundary effect. In the configuration of Sec.~2 the bulk
part cancels and $\Delta P_{\mathrm{mem}}$ is a genuine Casimir-type
correction.

Normalizing by the static fermionic Casimir pressure,
\begin{equation}
P_{0}=-\frac{7\pi^{2}}{960L^{4}},
\label{eq:static_pressure}
\end{equation}
we define
\begin{equation}
\mathcal{Q}_{P}(\xi)
\equiv
\frac{\Delta P_{\mathrm{mem}}}{H^{2}|P_{0}|}
=
\frac{960}{7\pi^{2}}\,\mathcal P_D(\xi),
\label{eq:pressure_response}
\end{equation}
so that, up to quadratic order in $H$,
\begin{equation}
P_{\mathrm{Cas}}
=
|P_{0}|
\left[
-1+H^{2}\mathcal{Q}_{P}(\xi)
\right].
\label{eq:total_pressure}
\end{equation}
A positive $\mathcal{Q}_{P}$ reduces the magnitude of the attractive
force, while a negative value reinforces it. Since $H\ll1$, the memory
term remains perturbative and does not reverse the sign of the total
force within the controlled regime.

\subsection{Limiting regimes}

For $\xi\ll1$, every discrete term in Eq.~\eqref{PD} is suppressed by
$\exp[-\pi^2/(2\xi^2)]$, and the exterior stress dominates:
\begin{equation}
\mathcal{Q}_{P}(\xi)
\simeq
-\frac{40}{49\pi^{3}}\xi^{4},
\qquad
\xi\ll1.
\label{eq:pressure_adiabatic}
\end{equation}
A slow pulse cannot bridge the MIT gap, so the interior stays nearly
empty, while the pairs produced outside push the plates together. The
memory then reinforces the attraction.

\begin{figure}[t]
    \centering
    \includegraphics[width=0.9\columnwidth]
    {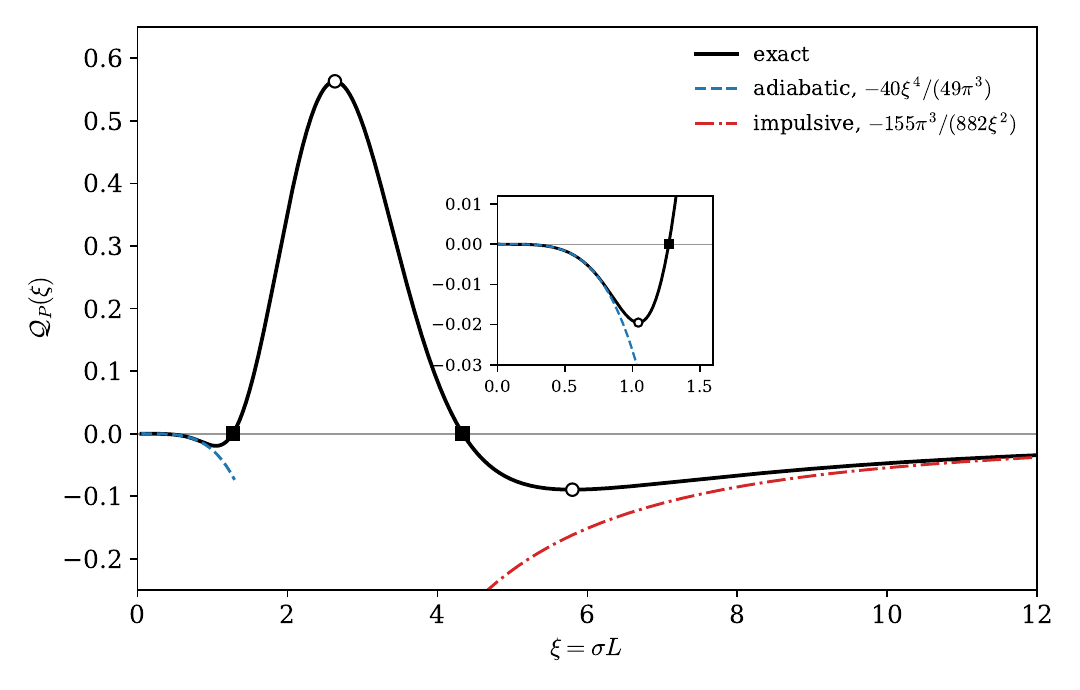}
    \caption{Normalized late-time memory contribution to the fermionic
    Casimir pressure,
    $\mathcal{Q}_{P}(\xi)=\Delta P_{\mathrm{mem}}/(H^{2}|P_{0}|)$, obtained
    from the difference between the interior and exterior normal stresses
    of the produced pairs at fixed occupation numbers, as a function of
    $\xi=\sigma L=L/\tau$. The solid black curve is the exact result,
    Eqs.~\eqref{PD} and \eqref{eq:pressure_response}; the dashed blue and
    dash-dotted red curves are the adiabatic and impulsive approximations,
    Eqs.~\eqref{eq:pressure_adiabatic} and \eqref{eq:pressure_impulsive}.
    Squares mark the zeros at $\xi\simeq1.27011$ and $\xi\simeq4.33735$;
    circles mark the extrema. The inset enlarges the adiabatic region,
    where $\mathcal{Q}_{P}\propto-\xi^{4}$.}
    \label{fig:pressure_response}
\end{figure}

For $\xi\gg1$, the half-integer Euler--Maclaurin formula~\eqref{EMhalf}
applies to $f(x)=(\pi x)^2J(\pi x;\xi)$. Near $b=0$,
\begin{equation}
 b^2J(b;\xi)=J(0;\xi)\,b^2-\frac43b^5
 +\sqrt{2\pi}\,\frac{b^6}{\xi}+O\!\left(\frac{b^7}{\xi^2}\right),
\end{equation}
so the first and third derivatives vanish at the origin, while
$f^{(5)}(0)=-160\pi^5$. With $B_6(1/2)=-31/1344$,
\begin{equation}
\mathcal{Q}_{P}(\xi)
\simeq
-\frac{155\pi^{3}}{882\,\xi^{2}},
\qquad
\xi\gg1.
\label{eq:pressure_impulsive}
\end{equation}
The response therefore reinforces the attraction for both slow and fast
pulses and vanishes in both limits. Between them, it changes sign twice.
Numerically,
\begin{equation}
\xi_0^{(1)}\simeq1.27011,
\qquad
\xi_0^{(2)}\simeq4.33735,
\end{equation}
with a positive maximum
$\mathcal{Q}_{P}^{\mathrm{max}}\simeq0.563174$ at
$\xi\simeq2.63196$, where the memory reduces the attraction, and negative
minima $\mathcal{Q}_{P}\simeq-0.0194932$ at $\xi\simeq1.04446$ and
$\mathcal{Q}_{P}\simeq-0.0895249$ at $\xi\simeq5.80002$. In the window
$\xi_0^{(1)}<\xi<\xi_0^{(2)}$ the lowest MIT levels are within the pulse
bandwidth, and the discrete interior spectrum, which concentrates
longitudinal momentum at $k_{z,n}\geq\pi/(2L)$, produces a larger normal
stress than the free continuum. Figure~\ref{fig:pressure_response} shows
the exact response together with its asymptotic approximations. We keep
$L$ fixed and vary $\sigma$, so that $\xi=\sigma L=L/\tau$ compares the
cavity's spectral timescale with the pulse duration.

In the present calculation the late-time remnant is carried by real fermion--antifermion pairs and is obtained from $|\beta|^2$. The same Bogoliubov coefficients determine vacuum persistence and the imaginary part of the in--out effective action. The response is dissipative in the external-field sense: the absorptive part describes energy transfer from the prescribed geometry to on-shell pairs. The underlying quantum evolution remains unitary, and no fundamental thermodynamic irreversibility is implied. In particular, a suitably prepared time-reversed process is not forbidden; what remains after the specified pulse is a non-vacuum out state. This interpretation assumes neither thermalization nor decoherence: in the free theory, the produced state is a coherent fermionic multipair state.

\FloatBarrier

\section{Conclusion}

We have obtained the leading boundary-dependent energy generated when a static MIT bag cavity is subjected to a weak, spatially homogeneous anisotropic gravitational pulse. Unlike the bosonic vacuum-polarization memory previously obtained for confined scalar and electromagnetic fields, the present effect is carried by real fermion--antifermion pairs and represents a boundary-dependent deposited energy. Three ingredients further distinguish the result. First, the shear couples through a first-order spinor vertex rather than a scalar second-order operator. Second, the MIT spectrum is half-integer and has no zero mode. Third, the Casimir subtraction can reverse the sign of the boundary-dependent response even though the pair-production energies in the confined and unbounded geometries are separately positive.

We have also derived the late-time force on the plates from the normal stress of the produced pairs at fixed occupation numbers, with the field and the pulse present on both sides of each plate. This force is not the $L$-derivative of the bulk-subtracted energy. It results from the imbalance between the interior stress of the discrete MIT spectrum and the exterior stress of the free continuum, whose produced-pair gas is anisotropic, with $p_z=5\epsilon/21$. The correction reinforces the Casimir attraction for slow and for fast pulses, as $-\xi^4$ and $-\xi^{-2}$ respectively, and reduces it in the window $1.27\lesssim\sigma L\lesssim4.34$.

The calculation is fully $(3+1)$ dimensional: the directions parallel to the plates remain continuous and only the normal momentum is discretized by physical boundaries. No compactification or lower-dimensional reduction is invoked. For $N_f$ independent massless Dirac species, both terms in Eq.~\eqref{totalenergy} are simply multiplied by $N_f$.

Our analysis is perturbative in the metric amplitude. The pair amplitude is required only to first order in $H$ to determine the deposited energy at order $H^2$; quadratic contact terms in the Dirac operator do not contribute at this order to the one-pair probability. Odd powers in the boundary-dependent energy are excluded by the symmetry $H\to-H$ combined with $x\leftrightarrow y$, which leaves the isotropic parallel-plate cavity unchanged. A massive extension is possible, but the MIT spectrum is then determined implicitly and the response becomes a two-parameter function of $mL$ and $\sigma L$. The massless problem already isolates the genuinely fermionic effect and provides a closed analytic response.

Because the effect is quadratic in the weak pulse amplitude, it is
expected to be extremely small for gravitational perturbations of
realistic magnitude. We therefore regard the present result primarily
as a theoretical probe of the interplay between gravitational particle
production and boundaries. Effective Dirac systems with externally
controlled anisotropic parameters may offer an analogue route, but a
concrete experimental mapping lies beyond the scope of this work.

The main results, Eqs.~\eqref{mainresult} and \eqref{Pmem}, identify a dissipative fermionic Casimir memory controlled by the competition between the pulse bandwidth and the bag gap. It provides a boundary-field-theory realization of gravitational fermion production while remaining physically and conceptually distinct from a static vacuum-energy shift.

The same Bogoliubov structure also implies particle--antiparticle mode
entanglement. In the out basis, each canonical fermionic channel has
the schematic form
$\alpha_j|0_j\bar 0_j\rangle+\beta_j|1_j\bar 1_j\rangle$ and is
therefore nonseparable whenever $0<|\beta_j|^2<1$.
This statement concerns mode entanglement between the particle and
antiparticle sectors; entanglement specifically between their spin
degrees of freedom depends on the rank and phase structure of the
matrix $\beta_{ss'}$ and is not evaluated here. Such correlations and
their boundary-dependent entropy constitute natural extensions of the
present work~\cite{Fuentes2010,Moradi2014}.

\section*{Acknowledgments}

C.R.M. acknowledges partial financial support from the Conselho Nacional de Desenvolvimento Cient\'ifico e Tecnol\'ogico (CNPq), through Grant No.~301122/2025-3. H.F.S.M. acknowledges partial financial support from CNPq under Grant No.~308049/2023-3.

\appendix
\renewcommand{\thesection}{Appendix \Alph{section}}

\section{Spin trace and normalization}
\label{app:spin}

For completeness, we give the short trace calculation underlying Eq.~\eqref{spinsum}. Let
\begin{equation}
 a^\mu=(0,k_x,-k_y,0),\qquad \mathcal W=\slashed a,
\end{equation}
and take $p^\mu=(E,\bm k,k_z)$ and $q^\mu=(E,-\bm k,-k_z)$. With covariantly normalized massless spinors,
\begin{equation}
 \sum_su_s(p)\bar u_s(p)=\slashed p,
 \qquad
 \sum_{s'}v_{s'}(q)\bar v_{s'}(q)=\slashed q.
 \label{completeness}
\end{equation}
The standard four-gamma trace then yields
\begin{align}
 \sum_{s,s'}|\bar u_s(p)\mathcal Wv_{s'}(q)|^2
 &=\Tr(\slashed p\slashed a\slashed q\slashed a)
 \nonumber\\
 &=8\left[E^2k_\perp^2-(k_x^2-k_y^2)^2\right].
 \label{apptrace}
\end{align}
The cavity modes entering Eq.~\eqref{beta} have unit one-particle norm. Relative to Eq.~\eqref{completeness}, each external spinor therefore supplies a factor $(2E)^{-1/2}$. Combining the resulting $(2E)^{-2}$ in $|\beta|^2$ with the pair energy $2E$ and the interaction factor $1/2$ gives precisely Eq.~\eqref{cavityenergy}; no additional spin-degeneracy factor is present. Finally, $k_x^2-k_y^2=k_\perp^2\cos(2\phi)$ and $\langle\cos^2(2\phi)\rangle_\phi=1/2$, which proves Eq.~\eqref{spinsum}.

\section{Longitudinal MIT-mode overlap}
\label{app:longitudinal_overlap}

In this appendix we verify explicitly that the spatially homogeneous shear
vertex does not mix distinct longitudinal MIT levels. For a fixed transverse
momentum $\mathbf{k}=(k_x,k_y)$, define
\begin{equation}
H_{\perp}
=
k_x\alpha^{\hat 1}
+
k_y\alpha^{\hat 2},
\qquad
\mathcal{O}_{\perp}
=
k_x\alpha^{\hat 1}
-
k_y\alpha^{\hat 2}.
\label{eq:transverse_operators}
\end{equation}
The stationary Dirac equation in the slab is
\begin{equation}
\left(
H_{\perp}
-i\alpha^{\hat 3}\partial_z
\right)\Psi(z)
=
\varepsilon\Psi(z).
\label{eq:stationary_dirac_slab}
\end{equation}
For a positive-energy mode with
\begin{equation}
E_{n\mathbf{k}}
=
\sqrt{k_\perp^2+k_{z,n}^2},
\end{equation}
the solution can be expressed in terms of its value at $z=0$ as
\begin{equation}
U_{ns}(z)
=
\left[
\cos(k_{z,n}z)
+
\frac{i}{k_{z,n}}
\alpha^{\hat 3}
\left(
E_{n\mathbf{k}}-H_{\perp}
\right)
\sin(k_{z,n}z)
\right]
U_{ns}(0).
\label{eq:positive_MIT_mode}
\end{equation}
The corresponding negative-energy branch entering the pair-creation
matrix element is
\begin{equation}
V_{ms'}(z)
=
\left[
\cos(k_{z,m}z)
-
\frac{i}{k_{z,m}}
\alpha^{\hat 3}
\left(
E_{m\mathbf{k}}+H_{\perp}
\right)
\sin(k_{z,m}z)
\right]
V_{ms'}(0).
\label{eq:negative_MIT_mode}
\end{equation}
The Fourier momentum appearing in Eq.~\eqref{eq:negative_MIT_mode} is the
same as that of the positive-energy mode. Because the negative-frequency
part of the field carries the opposite spatial phase, the physical
antiparticle momentum is $-\mathbf{k}$, as required by transverse
momentum conservation.

With the outward normals chosen at $z=0$ and $z=L$, the MIT conditions can
be represented by the complementary projectors
\begin{equation}
\Pi_0
=
\frac{1}{2}
\left(
1+i\gamma^{\hat 3}
\right),
\qquad
\Pi_L
=
\frac{1}{2}
\left(
1-i\gamma^{\hat 3}
\right),
\label{eq:MIT_projectors}
\end{equation}
up to the equivalent simultaneous reversal associated with the convention
for the normal. Imposing
\begin{equation}
\Pi_0\Psi(0)=0,
\qquad
\Pi_L\Psi(L)=0,
\end{equation}
on Eqs.~\eqref{eq:positive_MIT_mode} and
\eqref{eq:negative_MIT_mode} yields
\begin{equation}
\cos(k_{z,n}L)=0,
\qquad
k_{z,n}
=
\frac{\pi}{L}
\left(
n+\frac{1}{2}
\right),
\qquad
n=0,1,2,\ldots .
\label{eq:MIT_half_integer_spectrum_appendix}
\end{equation}

The longitudinal part of the pair-creation matrix element is
\begin{equation}
\mathcal{V}_{nm}^{ss'}
=
\int_0^L dz\,
U_{ns}^{\dagger}(z)
\mathcal{O}_{\perp}
V_{ms'}(z).
\label{eq:MIT_pair_matrix_element}
\end{equation}
The algebraic relations relevant to the boundary problem are
\begin{equation}
\left[\gamma^{\hat 3},\mathcal O_\perp\right]
=\left[\gamma^{\hat 3},H_\perp\right]=0,
\qquad
\left\{\alpha^{\hat 3},\mathcal O_\perp\right\}
=\left\{\alpha^{\hat 3},H_\perp\right\}=0.
\label{eq:MIT_vertex_algebra}
\end{equation}
The first relation shows that $\mathcal O_\perp$ and $H_\perp$ commute
with both MIT projectors,
\begin{equation}
\left[\Pi_{0,L},\mathcal O_\perp\right]
=
\left[\Pi_{0,L},H_\perp\right]
=0,
\label{eq:MIT_projector_vertex}
\end{equation}
and therefore map the MIT domain
$\mathcal D=\{\psi:\Pi_0\psi(0)=0,\ \Pi_L\psi(L)=0\}$ into itself.

\paragraph{Green boundary form.}
Let $\mathcal H_{\bm k}=H_\perp+K_z$, with $K_z=-\ii\alpha^{\hat3}\partial_z$.
Integration by parts gives
\begin{equation}
\langle\phi,\mathcal H_{\bm k}\chi\rangle
-\langle\mathcal H_{\bm k}\phi,\chi\rangle
=-\ii\left[\phi^\dagger\alpha^{\hat3}\chi\right]_0^L .
\label{eq:first_order_green}
\end{equation}
For $\phi,\chi\in\mathcal D$ the boundary term vanishes, because
$\phi^\dagger\alpha^{\hat3}\chi=\bar\phi\gamma^{\hat3}\chi$ is the normal
current, which the MIT conditions annul at each wall. Hence
$\mathcal H_{\bm k}$ is self-adjoint on $\mathcal D$. For the second-order
operator, applying Eq.~\eqref{eq:first_order_green} twice,
\begin{equation}
\langle\phi,\mathcal H_{\bm k}^2\chi\rangle
-\langle\mathcal H_{\bm k}^2\phi,\chi\rangle
=-\ii\left[\phi^\dagger\alpha^{\hat3}(\mathcal H_{\bm k}\chi)
+(\mathcal H_{\bm k}\phi)^\dagger\alpha^{\hat3}\chi\right]_0^L ,
\label{eq:second_order_green}
\end{equation}
which vanishes when $\phi$, $\chi$, $\mathcal H_{\bm k}\phi$ and
$\mathcal H_{\bm k}\chi$ all belong to $\mathcal D$. This is the condition
on the derivative boundary data that the projector argument alone does not
supply: through the first-order equation, it constrains $\partial_z\chi$
at the walls.

\paragraph{Application.}
Take $\phi=U_{ns}$ and $\chi=\mathcal O_\perp V_{ms'}$. Both $U_{ns}$ and
$\mathcal H_{\bm k}U_{ns}=E_{n\bm k}U_{ns}$ lie in $\mathcal D$. By
Eq.~\eqref{eq:MIT_projector_vertex}, $\mathcal O_\perp V_{ms'}\in\mathcal D$.
Moreover, using $\{K_z,\mathcal O_\perp\}=0$ and
$K_zV_{ms'}=-(E_{m\bm k}+H_\perp)V_{ms'}$,
\begin{equation}
\mathcal H_{\bm k}\mathcal O_\perp V_{ms'}
=H_\perp\mathcal O_\perp V_{ms'}
+\mathcal O_\perp(E_{m\bm k}+H_\perp)V_{ms'}\in\mathcal D,
\end{equation}
since every operator on the right-hand side commutes with the projectors.
The boundary form~\eqref{eq:second_order_green} therefore vanishes.
Finally, $\{H_\perp,K_z\}=0$ and $H_\perp^2=k_\perp^2$ give
$\mathcal H_{\bm k}^2=k_\perp^2+K_z^2$, and since $\mathcal O_\perp$
anticommutes with $K_z$, it commutes with $K_z^2$. Hence
\begin{equation}
\mathcal H_{\bm k}^2\,\mathcal O_\perp V_{ms'}
=\mathcal O_\perp\mathcal H_{\bm k}^2V_{ms'}
=E_{m\bm k}^2\,\mathcal O_\perp V_{ms'},
\end{equation}
and
\begin{equation}
\left(E_{n\bm k}^2-E_{m\bm k}^2\right)\mathcal V_{nm}^{ss'}=0.
\end{equation}
At fixed $\bm k$, $E_{n\bm k}^2\neq E_{m\bm k}^2$ for $n\neq m$, so
\begin{equation}
\mathcal V_{nm}^{ss'}
=
\delta_{nm}\mathcal V_n^{ss'}.
\label{eq:MIT_overlap_diagonal}
\end{equation}

\paragraph{Explicit modes.}
In terms of the explicit modes~\eqref{eq:positive_MIT_mode} and
\eqref{eq:negative_MIT_mode}, the matrix element contains
cosine--cosine, sine--sine and mixed sine--cosine integrals. For
half-integer momenta the mixed integrals do not vanish individually; for
instance,
$\int_0^L\dd z\,\cos(k_{z,0}z)\sin(k_{z,1}z)=L/\pi$. Their contributions vanish only after the spinor coefficients are included. Indeed, substituting
Eqs.~\eqref{eq:positive_MIT_mode} and
\eqref{eq:negative_MIT_mode} into
Eq.~\eqref{eq:MIT_pair_matrix_element}, the coefficients of the two
mixed integrals are
\begin{align}
C_{nm}^{(cs)}
&=-\frac{\ii}{k_{z,m}}
U_{ns}^{\dagger}(0)\,
\mathcal O_\perp\alpha^{\hat3}
\left(E_{m\bm k}+H_\perp\right)V_{ms'}(0),
\nonumber\\
C_{nm}^{(sc)}
&=-\frac{\ii}{k_{z,n}}
U_{ns}^{\dagger}(0)\,
\alpha^{\hat3}\left(E_{n\bm k}+H_\perp\right)
\mathcal O_\perp V_{ms'}(0),
\label{eq:mixed_spinor_coefficients}
\end{align}
where $(cs)$ and $(sc)$ multiply, respectively,
$\int_0^L\dd z\,\cos(k_{z,n}z)\sin(k_{z,m}z)$ and
$\int_0^L\dd z\,\sin(k_{z,n}z)\cos(k_{z,m}z)$. At $z=0$ the MIT
conditions imply
$U_{ns}(0)=\Pi_LU_{ns}(0)$ and
$V_{ms'}(0)=\Pi_LV_{ms'}(0)$. Since
$\mathcal O_\perp$ and $H_\perp$ commute with $\gamma^{\hat3}$,
whereas $\alpha^{\hat3}$ anticommutes with it, each operator between the
boundary spinors in Eq.~\eqref{eq:mixed_spinor_coefficients} exchanges
the two complementary MIT subspaces. Consequently,
\begin{equation}
\Pi_L\mathcal O_\perp\alpha^{\hat3}
\left(E_{m\bm k}+H_\perp\right)\Pi_L
=
\Pi_L\alpha^{\hat3}\left(E_{n\bm k}+H_\perp\right)
\mathcal O_\perp\Pi_L=0,
\label{eq:mixed_spinor_cancellation}
\end{equation}
and therefore $C_{nm}^{(cs)}=C_{nm}^{(sc)}=0$. Thus the nonvanishing
mixed trigonometric integrals make no contribution, while for $n\ne m$
the cosine--cosine and sine--sine integrals vanish by ordinary
orthogonality. This gives an explicit mode-by-mode verification of
Eq.~\eqref{eq:MIT_overlap_diagonal}. We have also verified this result numerically by
constructing the MIT modes in the Dirac representation, orthonormalizing
the two degenerate spin states of each level, and evaluating
$\mathcal V_{nm}^{ss'}$ by quadrature: the off-diagonal entries vanish to
machine precision.

Thus, although the plates break continuous translational invariance in the
normal direction, the homogeneous shear vertex does not generate
transitions between different half-integer MIT levels. The pair-production
probability consequently contains a single sum over $n$, rather than an
independent double sum over $n$ and $m$.

For modes normalized to unit one-particle norm, the explicit spin sum at a
fixed longitudinal level is
\begin{equation}
\sum_{s,s'}
\left|
\mathcal{V}_{n}^{ss'}
\right|^2
=
2k_\perp^2
-
\frac{
2\left(k_x^2-k_y^2\right)^2
}{
E_{n\mathbf{k}}^2
}.
\label{eq:MIT_mode_spin_sum}
\end{equation}
This agrees with the covariant trace calculation after the external
normalization factors are included, and we have confirmed it numerically
with the explicit modes. Averaging over the azimuthal angle,
\begin{equation}
\left\langle
\left(k_x^2-k_y^2\right)^2
\right\rangle_{\varphi}
=
\frac{k_\perp^4}{2},
\end{equation}
gives
\begin{equation}
\left\langle
\sum_{s,s'}
\left|
\mathcal{V}_{n}^{ss'}
\right|^2
\right\rangle_{\varphi}
=
2k_\perp^2
-
\frac{k_\perp^4}{E_{n\mathbf{k}}^2}.
\label{eq:angular_MIT_spin_sum}
\end{equation}
Combining Eq.~\eqref{eq:angular_MIT_spin_sum} with the pair energy
$2E_{n\mathbf{k}}$, the interaction factor $1/2$, and the Gaussian Fourier
profile reproduces the deposited energy used in the main text,
\begin{equation}
\frac{\Delta E_D(L)}{A}
=
\frac{\pi H^2}{\sigma^2}
\sum_{n=0}^{\infty}
\int\frac{d^2k}{(2\pi)^2}
e^{-2E_{n\mathbf{k}}^2/\sigma^2}
\left(
\frac{k_\perp^4}{2E_{n\mathbf{k}}}
+
\frac{k_{z,n}^2k_\perp^2}{E_{n\mathbf{k}}}
\right).
\label{eq:deposited_energy_overlap_check}
\end{equation}

\bibliographystyle{JHEP}
\bibliography{fermionic_casimir_memory}

\end{document}